# Magnetic and Structural properties of $MnV_2O_4$

Prashant Shahi[1], Saurabh Kumar[1], Neetika Sharma[2], Ripandeep Singh[2], P. U. M. Sastry[2], A. Das[2], A. Kumar[1], K. K. Shukla[1], A. K. Ghosh[3], A. K. Nigam[4], and Sandip Chatterjee[1,*]

[1] *Department of Applied Physics, Indian Institute of Technology (Banaras Hindu University), Varanasi-221005, India*

[2] *Solid State Physics Division, Bhabha Atomic Research Center, Anushakti Nagar, Mumbai, India*

[2] *Department of Physics, Banaras Hindu University, Varanasi-221005, India*

[3] *Department of CMP & MS, Tata Institute of Fundamental Research, Mumbai-400005, India*

ABSTRACT

Magnetization, neutron diffraction and X-ray diffraction of Zn doped $MnV_2O_4$ as a function of temperature have been measured and the critical exponents and magnetocaloric effect of this system have been estimated. It is observed, that with increase in Zn substitution the noncollinear orientation of Mn spins with the V spins decreases which effectively leads to the decrease of structural transition temperature more rapidly than Curie temperature. It has been shown that the obtained values of β, γ and δ from different methods match very well. These values do not belong to universal class and the values are in between the 3D Heisenberg model and mean field interaction model. The magnetization data follow the scaling equation and collapse into two branches indicating that the calculated critical exponents and critical temperature are unambiguous and intrinsic to the system. The observed double peaks in magneto-caloric curve of $Mn_{0.95}Zn_{0.05}V_2O_4$ is due to the strong distortion of $VO_6$ octahedra.

INTRODUCTION

In recent years the Vanadium oxide spinels have attracted much attention because of the orbital degeneracy, and the interplay of spin, orbital and lattice degrees of freedom. This interplay arises not only from conventional spin-orbit coupling but also from the fact that the occupation of the specific orbitals with geometrical anisotropy prefers specific types of magnetic interaction in specific direction (the so-called Kugel-Khomskii type coupling).[1] In spinel Vanadate the magnetic $V^{3+}$ ions with $t_{2g}$ orbital degeneracy are located at vertices of a network of corner sharing tetrahedral that is magnetically frustrating. Many studies have been done to understand the mechanism of these $AV_2O_4$ vanadates,[2-4] where A is a non-magnetic species such as Mg, Zn, or Cd. Usually Vanadates undergo a contraction along the c-axis, which favors the $d_{xy}$ orbital to be occupied by a $t_{2g}$ electron of every $V^{3+}$ ions. The second $t_{2g}$ electron either occupies different $d_{yz}$ and $d_{zx}$ orbitals alternately along the c-axis,[5] occupies the same $d_{yz} \pm id_{zx}$,[6] or forms the more complex pattern.[7] A common feature found in these materials is a sequence of two phase transitions.[2-4] The higher transition is a structural distortion involving a compression of the $VO_6$ octahedra and a consequent parallel lifting of the orbital degeneracy. The orbital ordering is accompanied, at lower temperature by an antiferromagnetic ordering.

Recent attention has turned to $MnV_2O_4$.[8-11] Here $Mn^{2+}$ is in the $3d^5$ high-spin configuration with no orbital degrees of freedom, and can be regarded as a simple S=5/2 spin. On the other hand, the B-site is occupied by the $V^{3+}$ ion, which takes the $3d^2$ high-spin configuration in the triply degenerate $t_{2g}$ orbital, and has orbital degrees of freedom. $MnV_2O_4$ exhibits a collinear ferrimagnetic ordering at $T_N$=57K, where the magnetic moments of the Mn and the V sites align to the opposite direction, and then structural phase transition from a cubic to a tetragonal phase at $T_s$=53K, with the spin structure becoming non-collinear.[8] It was also observed that the cubic to tetragonal transition could be induced by a magnetic field of few tesla.[9,10] In addition it has been pointed out[12] that the orbital state of $MnV_2O_4$ cannot be explained simply by anti-ferro orbital model. Moreover, in this compound large magneto-caloric effect is observed which is suggested to be related to the orbital entropy change due to the change of the orbital state of $V^{3+}$ ions with an applied field around $T_c$ (=57K).[13] Furthermore, when Zn is doped on the Mn site the value of magneto-caloric effect increases and the maximum is observed at the lower transition temperature ($T_{ST}$) which has been attributed to orbital ordering.[14] Zn doping also revealed that the structural phase transition from a cubic to a tetragonal phase is a co-operative phenomena dominated by orbital degrees of freedom.[9] In this perspective the detail study of Zn doping in $MnV_2O_4$ will be highly interesting to throw light on the mechamism of $MnV_2O_4$. Moreover, like distinguishing transport and magnetic behaviours, the critical behaviour of $MnV_2O_4$ might also be distinctive to provide interesting information about the magnetic spin ordering in this spinel Vanadate. Baek et al.[11] have reported the critical exponents of this system. From the ac-susceptibility measurement they obtained very unusal values of the critical constants ($\beta$ =0.36 and $\gamma$=0.59).[11] As a matter of fact, we have also studied the variation of critical constants with Zn doping.

EXPERIMENTAL

The polycrystalline $Mn_{1-x}Zn_xV_2O_4$ samples used in this study were prepared by solid state reaction method. Appropriate ratio of MnO and $V_2O_3$ were ground thoroughly and pressed into pellets. The pellets were sealed in evacuated quartz tube and heated at 950°C for 40 hours. The X-ray powder diffraction has been taken from Rigaku MiniFlex II DEXTOP X-ray Diffractometer with Cu-k$\alpha$ radiation. Magnetic measurement was done using MPMS SQUID (Quantum Design) magnetometer with the bulk samples. Data were collected upon warming up the sample. The Neutron diffraction measurement was performed on the neutron powder diffractometer (PD2) at Bhabha Atomic Research Centre, Mumbai, India using neutrons of wavelength, $\lambda$=1.2443Å.

RESULTS & DISCUSSION

Figure 1 shows the temperature dependence of magnetization of $Mn_{1-x}Zn_xV_2O_4$ under zero field cooled (ZFC) and field cooled (FC) condition at 100 Oe. The M-T curve of $MnV_2O_4$ exhibits a sharp paramagnetic-ferrimagnetic (PM-FM) phase transition. The Curie temperature $T_c$, defined by the minimum in dM/dT as a function of T, has been determined to be~58K. With further decrease of temperature the ZFC and FC curves exhibit a considerable divergence indicating the presence of frustration. It has been studied that the PM-FM transition as the second order transition and the transition at the lower temperature (~52K), $T_{ST}$ which is associated with the orbital ordering of V ions (also the structural transition from cubic to tetragonal phase) is the first order in nature.[8] Both the $T_C$ and $T_{ST}$ have been determined from the dM/dT vs. T curve (shown in the inset of Fig.1). For Zn=0.05 and 0.1 the $T_C$ decreased from 58K (x=0) to 52.5K and 47K respectively whereas the $T_{ST}$ decreased from 53K (for x=0) to 39K and 29K respectively. The decrease of $T_C$ can be explained as with the increase of Zn content the numbers of Mn-spins decrease. It is to be noted here that the size of the magnetization increases with the Zn content which is also consistent with the report by Adachi et al.[9] which is due to the change of canting angle with Zn doping.[9] Although both the $T_C$ and $T_{ST}$ decrease with Zn content but the gap between these two increases with increase of Zn concentration. It has been shown by Garlea et al.[15] that the structural transition is associated with the non-collinear ferrimagnetic structure (Mn spins are non-linearly oriented with the V spins). As Zn is non-magnetic, with increase of Zn the non-collinear orientation will decrease. The competition between these two may decrease the $T_{ST}$ more rapidly than $T_C$.

In order to monitor the changes in the crystal and magnetic structure across the transitions we have performed the Neutron diffraction and X-ray diffraction experiments. Figure 2 shows the evolution with temperature of (220) peak integrated intensities obtained from X-ray diffraction. It is observed from the integrated intensity vs. temperature curve that for all the three samples the structural transitions from cubic to tetragonal occurred. For x=0, 0.05 and 0.1 the $T_{ST}$ are, respectively, 50K, 40K and 32K. These values are consistent with those obtained from magnetic measurement. The transition can also be shown from the indexing of peaks at different temperatures. We have shown the splitting of peaks with temperature for x=0, 0.05 and 0.1 samples in Fig.3(a) which clearly shows the occurrence of tetragonal distortion by the splitting of the (400) peak into $(220)_T$ and $(004)_T$ peaks. In figure 3(b) the systematic occurrence of tetragonal distortion of $MnV_2O_4$ has been shown. Figure 4 shows a representative neutron diffraction pattern of $MnV_2O_4$ at 6K. We have analysed the diffraction pattern of all the samples with Reitveld refinement above and below the structural transition. The space group used above the $T_{ST}$ is Fd-3m and $I4_1/a$ below the transition. The parameters obtained from combined refinement of Xray diffraction and neutron diffraction data are summarized in Table I. The V-O bond length in the low temperature phase for $MnV_2O_4$ obtained from the fitting is 2.030(3) Å along the c-axis and 2.033(2) Å within the plane, where as for x=0.05 the values are respectively, 2.014(3) Å and 2.027(1) Å and for x=0.1 the values are respectively, 2.022(3) and 2.024(1). For all the cases the V-O values indicate the $VO_6$ octahedral is compressed along the c-direction and with Zn concentration initially (x=0.05) the distortion increases and with further increase of Zn content (x=0.1) the distortion decreases. This distortion splits the triply degenerate $t_{2g}$ orbitals into an *xy* orbital with lower energy and doubly degenerate *yz* and *zx* orbital with a higher energy. It has been explained that in $MnV_2O_4$ Orbital ordering occurs immediately after the spin ordering because the long-range spin ordering endows the orbital system with a one-dimensional character along the c-axis which produces a strong Jahn-Teller lattice distortion coupled with Orbital ordering.[17] In the present system with doping of Zn the Jahn-Teller distortion increases for the Zn=0.05 concentration and for x=0 and x=0.1 the distortion are nearly equal. But as has been mentioned with increase of distortion the gap between $T_{ST}$ and $T_C$ increase which is inconsistent with the explanation by Zhou et al.[16] Therefore, as we have already discussed that the more rapid decrease of $T_{ST}$ is due to the fact that with Zn-doping the effective exchange interaction between Mn and V decrease which in effect decreases the temperature where non-collinear ferromagnetic state appears which leads to the rapid decrease of $T_{ST}$ with Zn doping.

It is to be mentioned that near the two phase transitions the variation of order parameter with temperature can be described as $I(T) \propto (T_{C,ST}-T)^{2\beta}$, where $\beta$ is the critical exponent. Garlea et al. determined the $\beta$ value from the integrated intensity I(T) fitting.[15] The obtained value by them was close to 3D Heisenberg and 3D Ising models. We have also estimated critical exponents for all the samples using modified Arrot plots.

According to the Scaling hypothesis,[17] a second order magnetic phase transition near the Curie point is characterized by a set of critical exponents of $\beta$, $\gamma$ and $\delta$ and the magnetic ordering can be studied

$$(H/M)^{1/\gamma} = C_1(T-T_c) + C_2 M^{1/\beta} \qquad (1)$$

which combines the relations for the spontaneous magnetization below $T_c$

$$M \sim (T_c - T),$$

and the inverse magnetic susceptibility above $T_c$

$$\chi^{-1} \sim (T-T_c)$$

To find the correct values of $\beta$ and $\gamma$, linear fits to the isotherms are made to get the intercepts giving M(T) and $\chi$(T). These new values of $\beta$ and $\gamma$ are then used to make a new modified Arrott plot. New values of critical exponents thus obtained are re-introduced in the scaling of the modified Arrott plot. The process is repeated until the iteration converges, leading to an optimum fitting value.

Figure 5 shows the final result for the $Mn_{1-x}Zn_xV_2O_4$ samples. We have taken M(H) isotherms from 50K to 70K in every 2K interval. The calculated values of $\beta$ and $\gamma$ are, respectively, 0.393 and 1.01 for x=0 sample. For x=0.05 the values are respectively, 0.40 and 1.02 whereas for x=0.1 the values of $\beta$ and $\gamma$ are respectively, 0.42 and 1.07. Below 54 K it deviates from linearity due to the first order transition at ~52K.

The critical values we obtained from different methods are not from the universality class. According to the scaling theory the magnetization equation can be written as $M(H,\varepsilon)\varepsilon^{-\beta} = f_{\pm}(H/\varepsilon^{\beta+\gamma})$, where $\varepsilon$ is the reduced temperature, $f_+$ for $T > T_c$ and $f_-$ for $T < T_c$ are regular functions.[18] The equation states that the plot between $M|\varepsilon|^{-\beta}$ vs $H|\varepsilon|^{-(\beta+\gamma)}$ gives two universal curves: one for $T > T_c$ and other for $T < T_c$. As shown in fig.6, the curves are divided in two parts one above $T_c$ and one below $T_c$. The inset of the fig.6 also shows log-log plot and this also falls into two classes one above $T_c$ and one below $T_c$, in agreement with the scaling theory. Therefore the FM behaviour around curie temperature get renormalized following the scaling equation of state indicating that the calculated critical exponents are reliable. Moreover, exponents often show various systematic trends or crossover phenomenon's one approaches $T_c$.[19,20] This occurs due to the presence of various competing couplings and/or disorder. For this reason, it is useful to introduce temperature-dependent effective exponents for $\varepsilon \neq 0$. It can be mentioned that effective exponents are non universal properties, and they are defined as:

$$\beta^{eff}(\varepsilon) = \frac{d[\ln M_S(\varepsilon)]}{d[\ln \varepsilon]} \quad . \quad \gamma^{eff}(\varepsilon) = \frac{d[\ln \chi_0^{-1}(\varepsilon)]}{d[\ln \varepsilon]} \tag{8}$$

We have calculated the $\beta^{eff}$ and $\gamma^{eff}$ by using the equation 8 (not shown) they do not match with universality class.

It is observed from the above discussion that the critical exponents of the present investigated sample are not consistent with the universality class. The similar behaviour has been observed in perovskite $Pr_{0.5}Sr_{0.5}MnO_3$ ($\beta = 0.397$ and $\gamma = 1.331$)[21] and in $La_{0.7}Sr_{0.3}MnO_3$ ($\beta = 0.45$ and $\gamma = 1.2$)[22] due to phase separation. Other than perovskite, $Gd_{80}Au_{20}$ also shows unusal critical exponents ($\beta = 0.44(2)$ and $\gamma = 1.29(5)$) arise due to the dilution of global spin with the substitution of non-magnetic ions.[23] In the present case very close to the second order PM-FM transition (at~58K) there exists a first order structural transition (~52K) which is associated with the collinear to non-colliner spin transition. This may cause a large spin fluctuation which may be responsible for the unusal critical exponents between the actual material and the theoretical model.

In order to further investigate we have also estimated the magneto caloric effect of all the samples. The magnetic entropy change is given by and can be evaluated by the expression

$$|\Delta S_m| = \sum_i \frac{M_i - M_{i+1}}{T_{i+1} - T_i} \Delta H_i \tag{7}$$

Where $M_i$ is the Magnetization at Temperature $T_i$.[24] The obtained $|\Delta S_m|$ has been plotted as a function of temperature in figure 7. In the inset of fig.7 the magnetic field variation of $|\Delta S_m|$ shows the $H^{2/3}$ dependency (the value of the exponent for x=0, 0.05 and 0.1 are respectively, 0.63, 0.609 and 0.61). This is consistent with the relation between magnetic entropy and the magnetic field near the magnetic phase transition which is given by[25]

$$|\Delta S_m| = -1.07qR(g\mu_B JH/kT_c)^{2/3} \tag{8}$$

where q is the number of magnetic ions, R is the gas constant, and g is the Landau factor. In fig.7 it is observed that as Zn content increases the magnetic entropy value decreases, which is also consistent with the expression 8. As with increase of Zn content the magnetic Mn ions decrease. Moreover, in $Mn_{1-x}Zn_xV_2O_4$ for x=0 and 0.1 only one peak is found and that is at $T_c$ but for x=0.05 two peaks are observed (one is at $T_c$ and another is at $T_{oo}$). The observed behaviour for x=0 and x=0.05 is consistent with those reported.[13,14] It has been explained that the large magneto caloric effect in $MnV_2O_4$ is due to the change of the orbital state of $V^{3+}$ ions with applied field around $T_c$ which leads to the change in orbital entropy.[13] The observed second peak in x=0.05 is suggested to be due to the strong coupling between orbital and spin degrees of freedom.[14] But in that case in $MnV_2O_4$ sample also we should get two peaks. Moreover, in the present investigation x=0.1 sample does not show second peak. It has already been discussed that maximum distortion of $VO_6$ octrahedral occurs for x=0.05 and for x=0.1 the distortion is nearly equal to the undoped one. Therefore, for x=0.05 the orbital ordering becomes maximum which in effect increase the coupling between orbital and spin degrees of freedom at $T_{oo}$ leading the second peak in x=0.05 sample. When Zn content increases (*viz.* x=0.1) it is observed that distortion of $VO_6$ octrahedral becomes nearly equal to the $MnV_2O_4$ system leading to the disappearance of second peak at $T_{oo}$.

CONCLUSION:

The decrease of $T_C$ with increase of Zn content in $Mn_{1-x}Zn_xV_2O_4$ can be explained as with the increase of Zn content the numbers of Mn-spins decrease. The size of the magnetization increases with the Zn content which is due to the change of canting angle with Zn doping. With increase of Zn the non-linear orientation of Mn spins with the V spins will decrease which effectively decrease the structural transition temperature more rapidly than Curie temperature. It has been shown that the obtained values of the critical exponents β, γ and δ do not belong to universal class and the values are in between the 3D Heisenberg model and mean field interaction model. The magnetization data follow the scaling equation and collapse into two branches indicating that the calculated critical exponents and critical temperature are unambiguous and intrinsic to the system. The observed double peaks in $Mn_{0.95}Zn_{0.05}V_2O_4$ is due the strong distortion of $VO_6$ octahedra. In this composition (x=0.05 ) the orbital ordering becomes maximum which in effect increase the coupling between orbital and spin degrees of freedom at $T_{oo}$ leading the second peak in magneto-caloric behavior in x=0.05 sample . When Zn content increases (*viz.* x=0.1) it is observed that distortion of $VO_6$ octrahedral becomes nearly equal to the $MnV_2O_4$ system leading to the disappearance of second peak at $T_{oo}$.

**ACKNOWLRDGEMENT:** SC is grateful to the funding agencies DST (Grant No.: SR/S2/CMP-26/2008) and CSIR (Grant No.: 03(1142)/09/EMR-II) for financial support. Authors are also grateful to D. Budhikot for his help in magnetization measurement.

**Figure Captions:**

1. Temperature variation of magnetization of $Mn_{1-x}Zn_xV_2O_4$ (with x=0, 0.05, 0.1) measured at H=100 Oe. Inset shows the plot of dM/dT vs. T indicating two transitions. (colour online)

2. Integrated intensity of (220) for $Mn_{1-x}Zn_xV_2O_4$ (with x=0, 0.05, 0.1) (colour online)

3. (a) Splitting of 440 → $(220)_T$ + $(004)_T$ of $Mn_{1-x}Zn_xV_2O_4$ (with x=0, 0.05, 0.1) obtained from XRD.(b) shows systematic occurrence of Tetragonal distortion for x=0.1 sample. (colour online)

4. A representative neutron diffraction pattern of $MnV_2O_4$ at 6K (colour online)

5. Final results for critical constants of $Mn_{1-x}Zn_xV_2O_4$ (with x=0, 0.05, 0.1). (colour online)

6. Universal curves and inset shows the log–log plot of universal curves of $Mn_{1-x}Zn_xV_2O_4$ (with x=0, 0.05, 0.1).

7. Magneto caloric effect and Inset shows fitting of ΔS vs H curve of $Mn_{1-x}Zn_xV_2O_4$ (with x=0, 0.05, 0.1).

Table 1. Structural parameters (lattice parameters, bond lengths) of $Mn_{1-x}Zn_xV_2O_4$ (with x=0, 0.05, 0.1) samples obtained from Reitveld refinement. The structural data have been refined with Space group I41/a at 6K and Fd-3 m at 300K.

|  | 6K | | | | 300K | | |
|---|---|---|---|---|---|---|---|
| Bonds (Å) | X=0.0 | X=0.05 | X=0.1 | Bonds (Å) | X=0.0 | X=0.05 | X=0.1 |
| MN- O | 4x 2.034(3) | 2x2.032(2) | 2.021(2) | MN- O | 4x 2.043(1) | 4x 2.0338(6) | 4x 2.0326(9) |
| MN- O |  | 2x2.032(2) | 2.021(2) |  |  |  |  |
| V- O | 2x 2.030(3) | 2x2.014(3) | 2.022(3) | V - O | 6x 2.023(1) | 6x 2.0249(6) | 6x 2.0234(9) |
| V - O | 4x2.033(2) | 4x2.027(1) | 2.024(1) | V- V | 3.0130 (1) | 6x 3.0105 (1) | 6x 3.0105 (1) |
| V- V | 4x3.0117(2) | 4x 3.0029(2) | 2.9990(2) |  |  |  |  |
| V- V |  | 2x 3.0167(2) | 3.0102(2) |  |  |  |  |
| a(Å) | 6.0555(3) | 6.0333(3) | 6.0203(4) |  | 8.5220671 | 8.5150299 | 8.5092974 |
| b(Å) | 6.0555(3) | 6.0333(3) | 6.0203(4) |  | 8.5220671 | 8.5150299 | 8.5092974 |
| c(Å) | 8.4726(9) | 8.4546(8) | 8.4507(9) |  | 8.5220671 | 8.5150299 | 8.5092974 |
| $M_{Mn}$ $M_V$ | X=0.0 at 6K 4.7 $\mu_B$ parallel to c. 0.5 $\mu_B$ at an angle with c. | | | X=0.05 at 6K 4.7 $\mu_B$ parallel to c . 0.5 $\mu_B$ at an angle with c . | | X=0.1 at 6K 4.7 $\mu_B$ parallel to c . 0.5 $\mu_B$ at an angle with c. | |

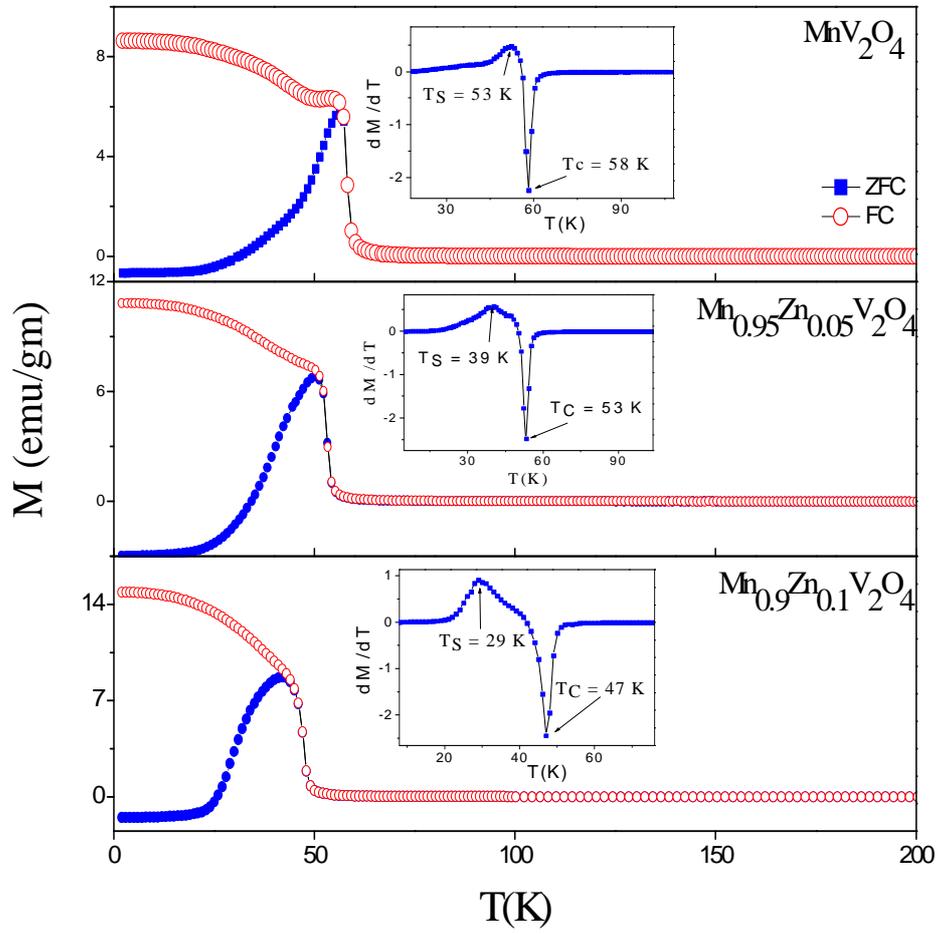

Figure 1

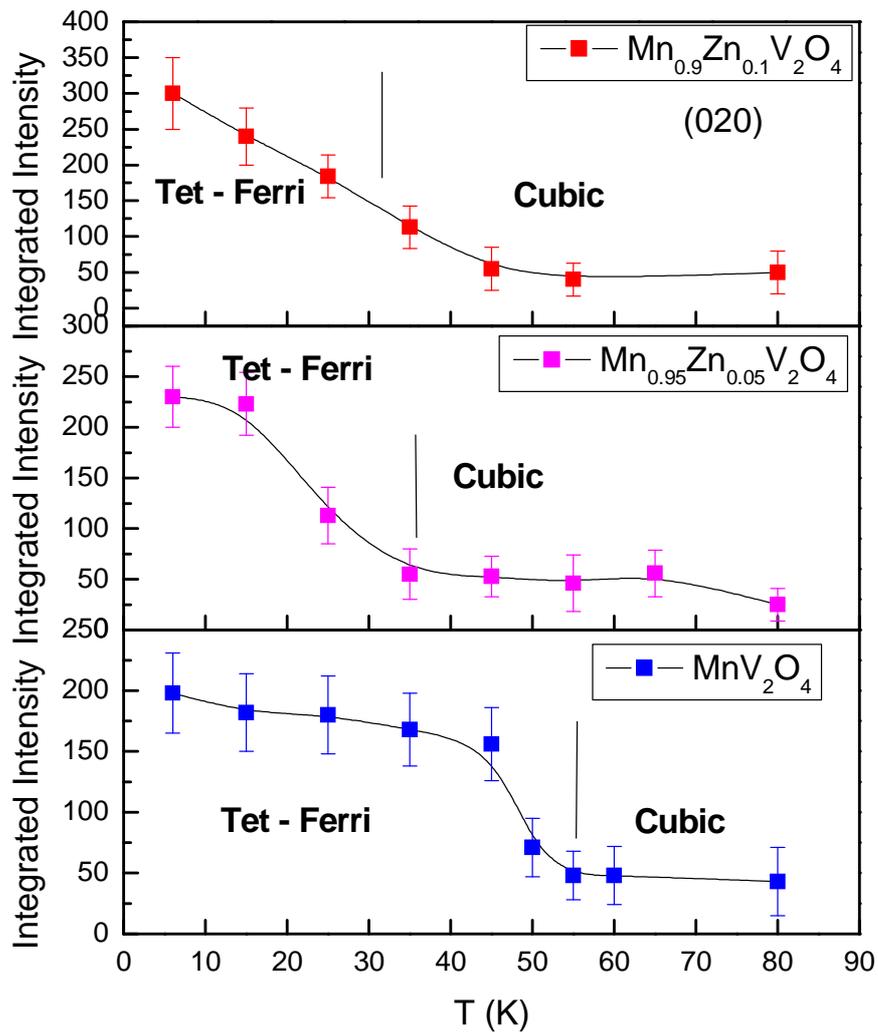

Figure 2

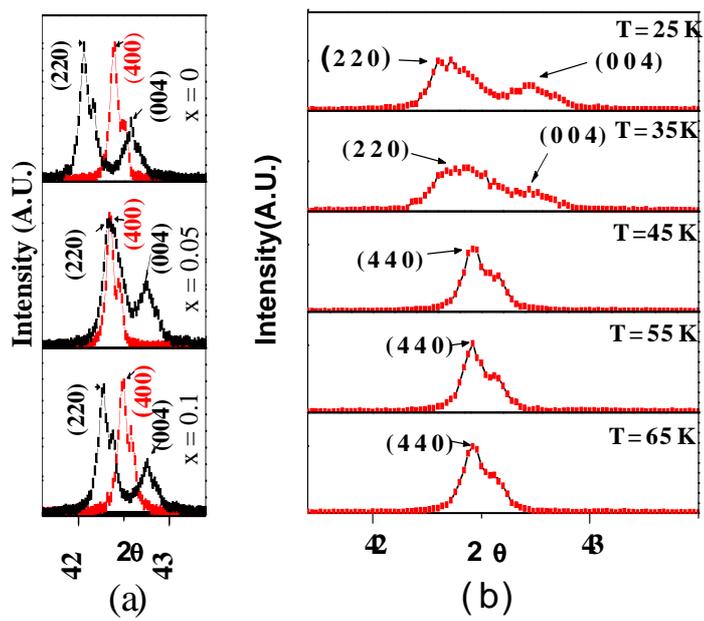

Figure 3

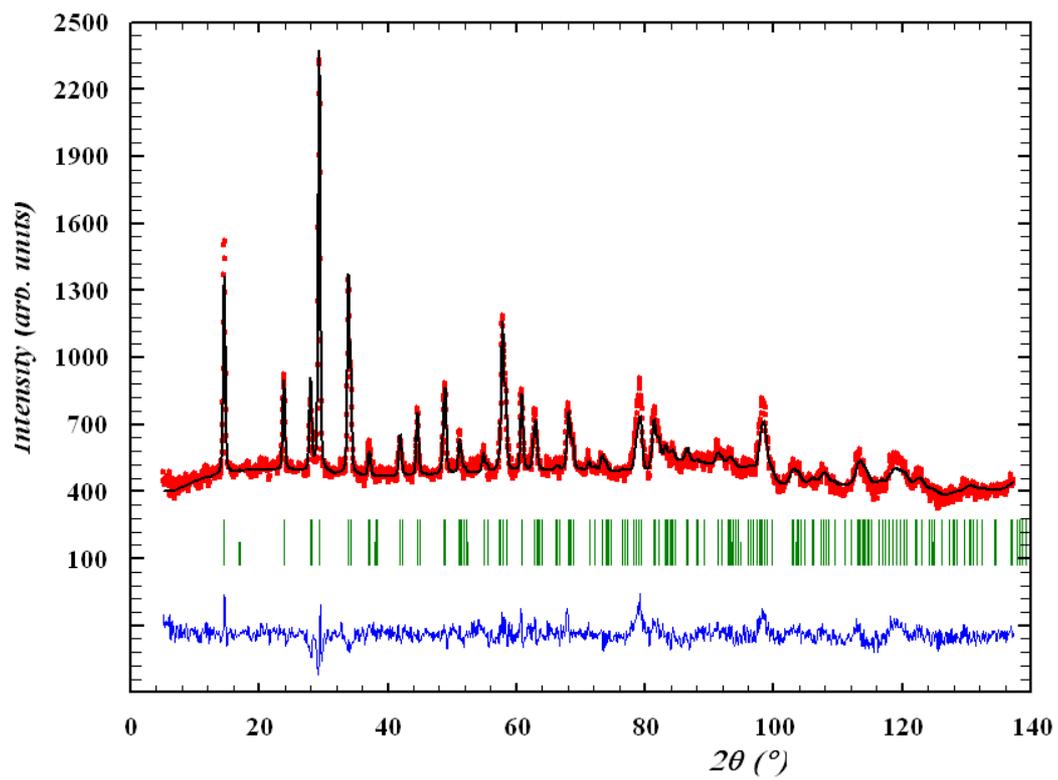

Figure 4

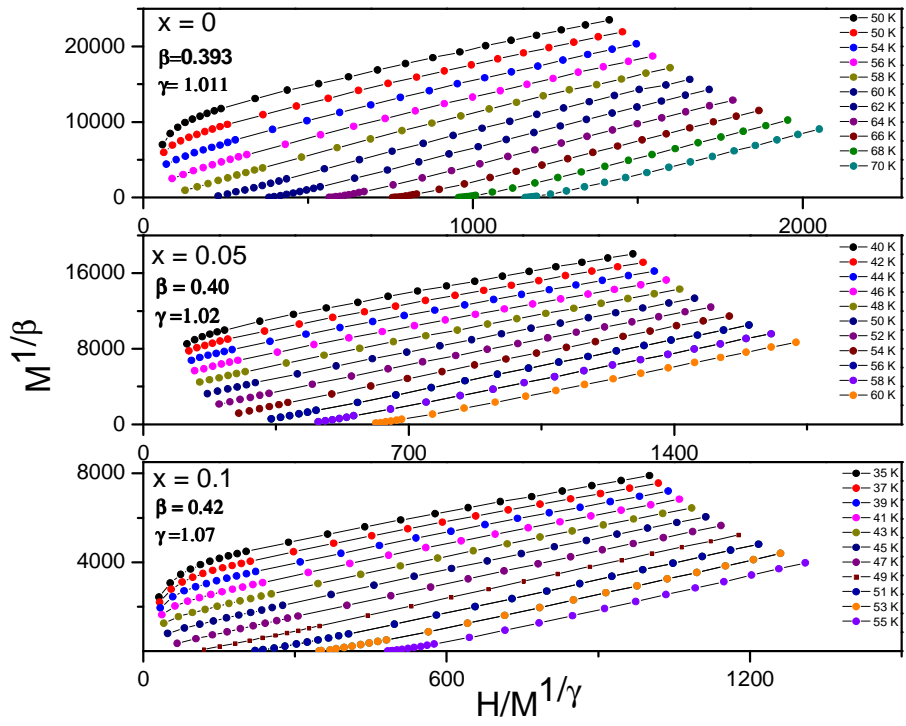

Figure 5

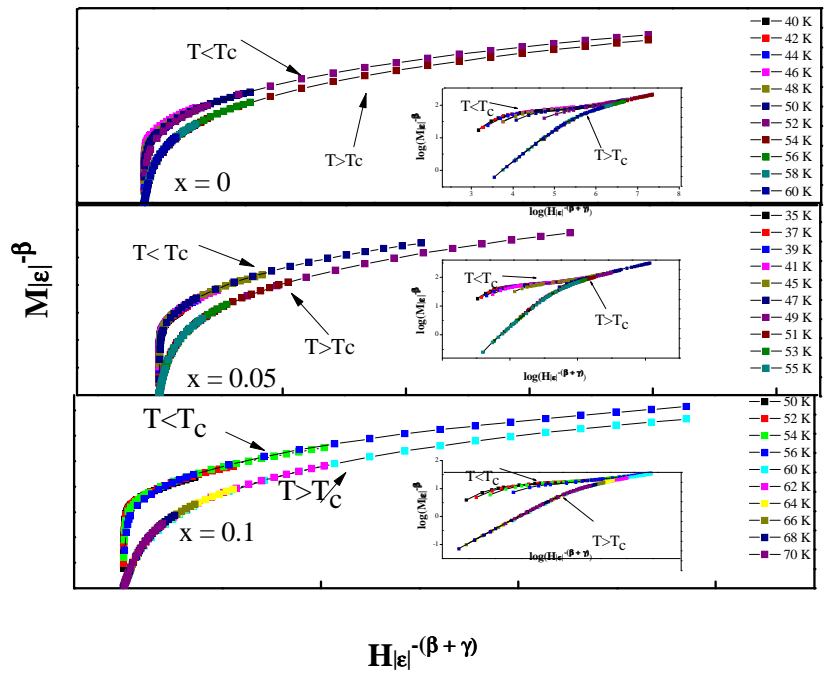

Figure 6

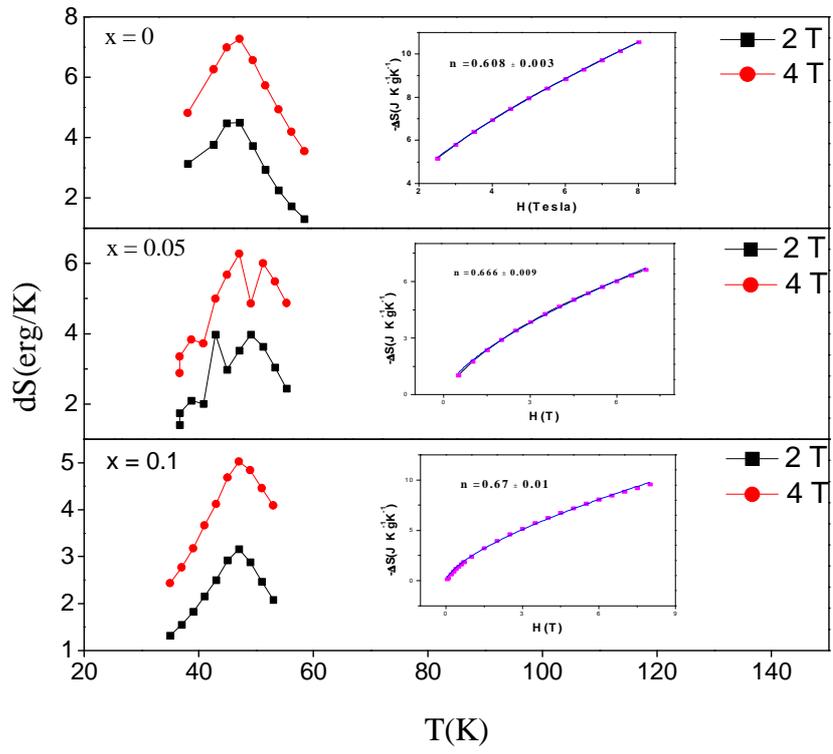

Figure 7